\documentclass[preprint,nofootinbib,aps,superscriptaddress,eqsecnum]{revtex4}
\usepackage{epsf}

\newcommand{\beq}{\begin{equation}}
\newcommand{\eeq}{\end{equation}}
\newcommand{\beqa}{\begin{eqnarray}}
\newcommand{\eeqa}{\end{eqnarray}}
\def\bea{\begin{eqnarray}}
\def\eea{\end{eqnarray}}

\def\lp{\lambda^\prime}

\def\ifmath#1{\relax\ifmmode #1\else $#1$\fi}

\def\pslash{p \hspace{-0.4em}/}
\def\qslash{q \hspace{-0.5em}/}
\def\ra{\rightarrow}

\def\lp{\lambda'}

\def\ie{ {\em i.e.}}

\begin{document}

\preprint{\vbox{
\hbox{}
\hbox{}
\hbox{}
\hbox{}
\hbox{TIFR/DHEP/INO-1201}
\hbox{December 2008} }}

\vspace*{3cm}

\title{Pion Spectra in the Production of Resonances by Neutrinos}

\author{E.~A.~Paschos}\email{paschos@physik.uni-dortmund.de}
\affiliation{Institut f\"{u}r Physik,Technische Universit\"{a}t
Dortmund, 
D-44221 Dortmund, Germany \vspace*{6pt}}

\author{Subhendu Rakshit}\email{rakshit@tifr.res.in}
\affiliation{Institut f\"{u}r Physik,Technische Universit\"{a}t
Dortmund, 
D-44221 Dortmund, Germany \vspace*{6pt}}
\affiliation{Department of High Energy Physics,
Tata Institute of Fundamental Research,
Homi Bhabha Road,
Mumbai -- 400 005,
India.  \vspace*{16pt}}

\begin{abstract}
A method is presented using helicity cross sections for calculating
neutrino-nucleon interactions. The formalism is applied in the
calculation of the pion spectra produced by ${\nu}_{\mu}$ and
$\nu_{\tau}$ beams. The masses of the charged leptons are kept
throughout the calculations. Cross sections are presented in numerous
figures where the contributions of the significant form factors are
also shown. The article describes the steps of the calculation and
gives details so that it can be reproduced and adapted to the
kinematic conditions of the experiments.

\end{abstract}

\maketitle

\section{Introduction}
\label{sec:introduction}

Neutrino production of resonances is attracting a lot of attention
because differential cross sections will be measured in a new
generation of experiments which will try to verify the functional form
of the cross sections
\ie\, the number and the $Q^2$ dependence of form factors. They
will also be used as an input to study properties of neutrinos in
oscillation experiments. The dominant signal at low neutrino energies
will be the $\Delta$-resonance. Many of the completed experiments
detected the energy and the angle of the produced muon which motivated
theoretical authors to integrate over the phase space of the decay
products, thus presenting cross sections $\frac{d\sigma}{dQ^2}$,
$\frac{d\sigma}{dW}$ and the integrated cross
section~\cite{PasSakuda,PPY,AlvarezRuso:2007tt,Graczyk:2007zz,athar}. The
comparisons of the calculated cross sections (differential or
integrated) are consistent with the data, but we must confess the
error bars are large so that there is a significant spread on the
experimental points. Theoretical calculations of the pion spectra are
also available~\cite{athar,PSU,siegen}.\\

The accuracy will improve in the new experiments and explicit
distributions on the energy spectrum of the pions produced in the
decays will become available. This motivates us to calculate the pion
spectrum from the diagram in Fig.~1 by keeping the $\Delta$-propagator
and without integrating over the whole phase space of the pion.  We
shall present the calculation in detail so that the interested reader
can reproduce and use our results. We will also make available a code
for our calculation which can be used by experimentalists.\\

The method that we adopt decomposes the leptonic tensor into helicity
components and uses helicity cross sections for the scattering of the
$W^\pm$ or $Z^0$ bosons on the nucleons. This method has been found to
be useful~\cite{BjPaschos,PaschosBook} and was adopted recently in the
coherent pion production by neutrinos~\cite{GKP}. This way the
calculation, whose algebra is long and tedious, simplifies. We decided
to present results for free protons and neutrons in order to show
their main features and separate them from nuclear target effects. We
also take the opportunity to mention and correct a mistake in the pion
spectrum that appears in an earlier article on which one of us (EAP)
was a co-author~\cite{PPY}.\\

 In addition to describing the formalism we use it for the calculation
 of the differential and integrated cross sections. We calculate the
 energy spectra of produced pions by $\nu_{\mu}$ and $\nu_{\tau}$
 beams. We keep the masses of the charged leptons throughout the
 calculations or set them equal to zero in order to see the changes
 brought about in the spectra. We also show in many figures the
 contributions of the important form factors and their interferences
 explicitly.  Section~\ref{sec:method} describes the method and
 includes detailed formulas for the cross sections. The functional
 dependence of the form factors are included in
 section~\ref{sec:Numerical}, where they are used in order to
 calculate the results in figures 2-10. A summary of the results and
 of the improvements that have taken place over the past few years are
 included in the last section.

\section{The Method}
\label{sec:method}

The process we consider is in general
\beq
\nu_{\ell}(\vec k)\, N(\vec p)\ra \ell^-({\vec k}^\prime) {\cal R}({\vec
p}_{\Delta})\ra\ell^-({\vec k}^\prime)\, \pi(p_{\pi})\,N(p^\prime)
\eeq
with ${\cal R}$ being a resonance with spin $3/2$.
\begin{figure}[htb]
\vspace*{1.5cm}
\centerline{ \epsfxsize=7.3cm\epsfysize=4.5cm
\epsfbox{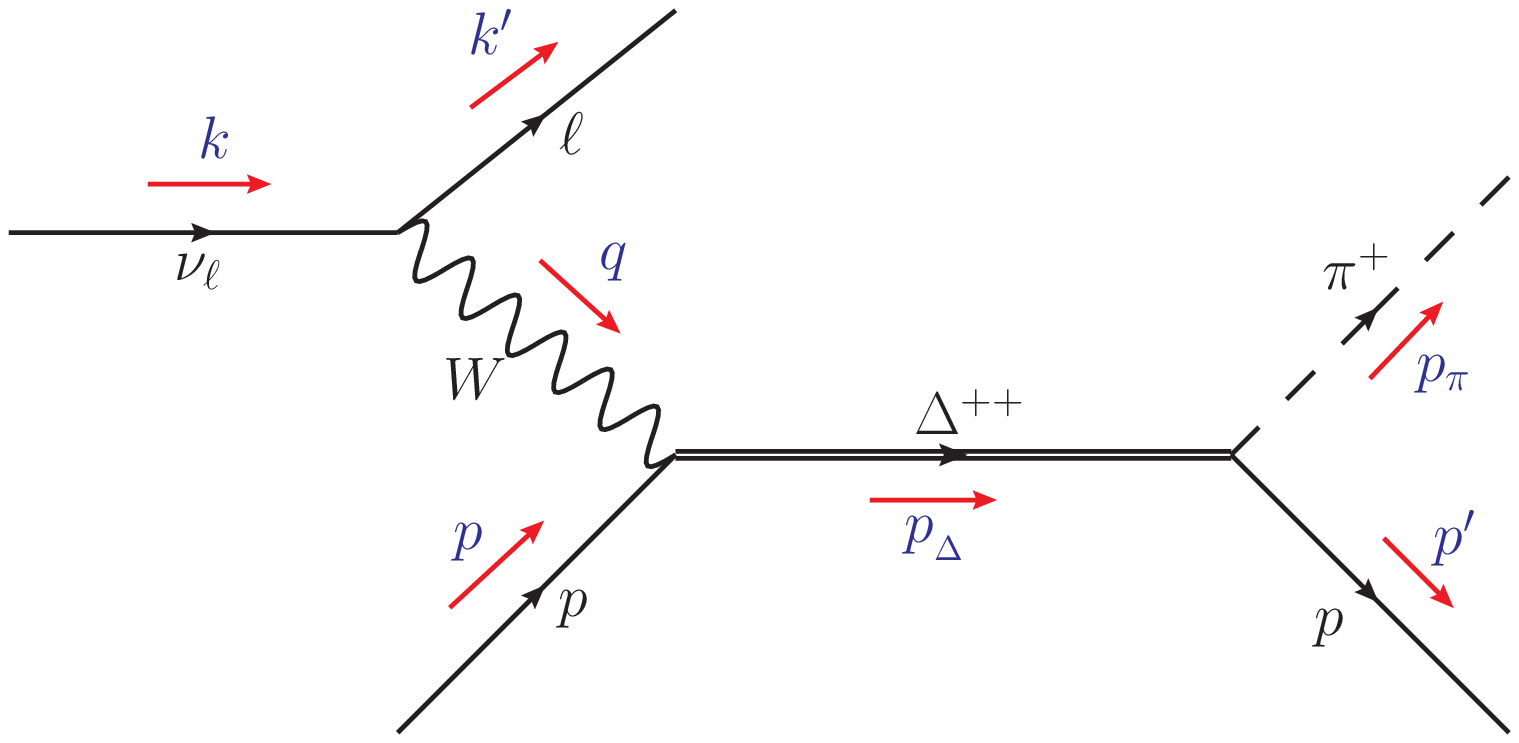}  
\hspace*{1cm}
 \epsfxsize=7.3cm\epsfysize=4.5cm
\epsfbox{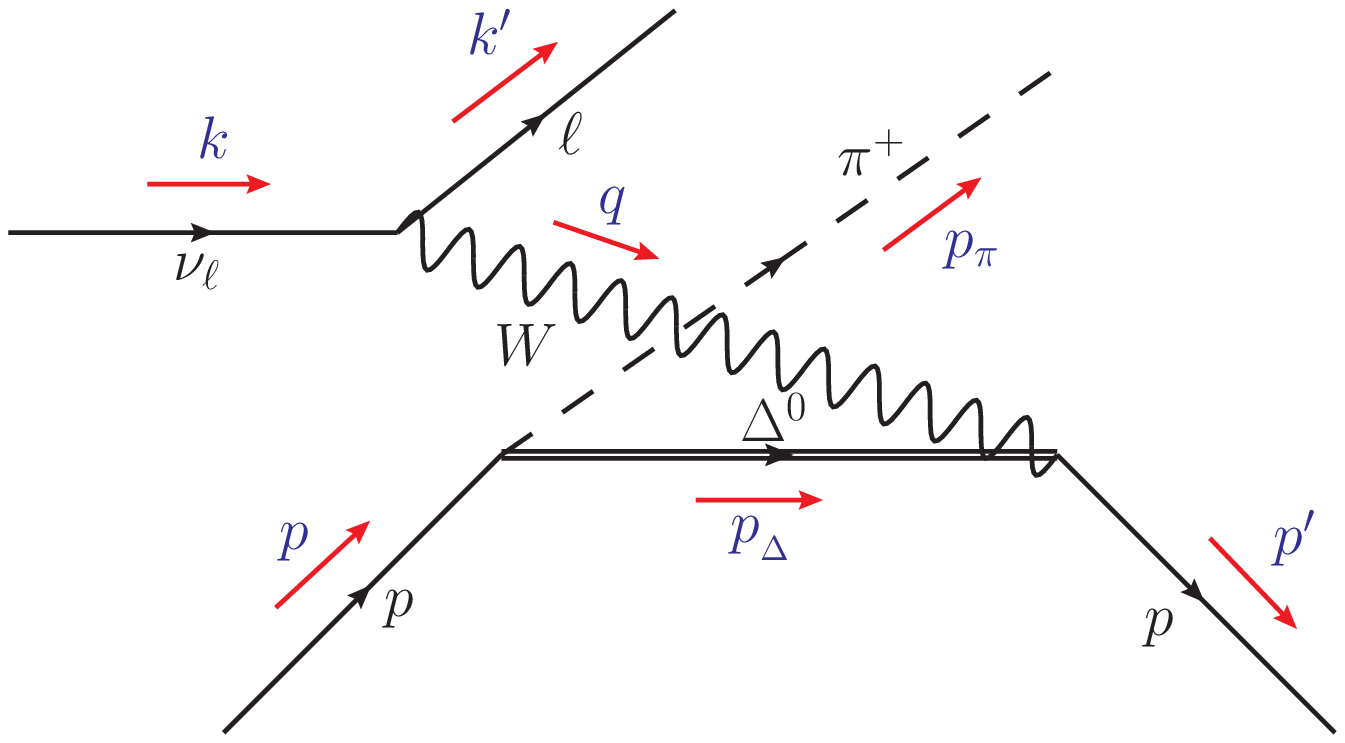} } 
\caption{\em $s$-channel and $u$-channel diagrams for the process 
$\nu_{\ell} p \ra \ell \pi^+ p$
         } \label{diagram}
\vspace*{-0.2cm}
\end{figure}
It is convenient to
use variables in the rest frame of the nucleon
\beq q = k-k^\prime,\quad Q^2 = -q^2,\quad W^2=p_{\Delta}^2,\quad
\nu=E-E^\prime
\eeq
inspired from the kinematics of deep inelastic scattering. The
leptonic tensor is
\beq
{\cal
L}_{\mu\nu}=4\left[k_{\mu}k'_{\nu}+k_{\nu}k'_{\mu}+g_{\mu\nu}
k.k'-i \epsilon_{\mu\nu\alpha\beta} k^{\alpha}k'^{\beta}\right] =
\sum_{h,h'} L_{h'h} \varepsilon_{h'}^{\mu *} \varepsilon_{h}^{\nu}
\label{dense}
\eeq
which can be decomposed in terms of the polarizations of the exchanged
current. When we keep the mass of the muon or tau lepton, there are
polarizations for the spin-1 and zero states. In the laboratory frame
we introduce the basis vectors
\bea
&&\varepsilon^{\mu}_R = \frac{1}{\sqrt 2}\left(0,1,i,0\right)\nonumber \\
&&\varepsilon^{\mu}_L = \frac{1}{\sqrt 2}\left(0,1,-i,0\right)\nonumber \\
&&\varepsilon^{\mu}_0 = \frac{1}{\sqrt {Q^2}}\left(|\vec
q|,0,0,q_0\right)
\eea
for helicities and the scalar component
\beq
\varepsilon^{\mu}_{\ell} = \frac{q^{\mu}}{\sqrt {Q^2}}.
\eeq 
We warn the reader that various definitions occur in the published
articles which differ from each other.  For instance, the above
notation is slightly different from than in ref.~\cite{GKP}. Numerous
articles in the early studies of resonance production and recently
\cite{Sehgal,kuzmin} define a set of polarizations in the rest frame
of the resonance because it simplifies calculations. The above set is
convenient with the first three polarizations being present even when
the leptons are massless and the longitudinal component appearing for
massive leptons. It is also a complete and orthonormal set of
polarizations.\\

The coefficients $L_{h'h}$ are obtained by inverting Eq.~\ref{dense}
\beq L_{h'h} ={\cal L}_{\mu\nu} \varepsilon^{\mu*}_{h'}
\varepsilon^{\nu}_h.
\eeq
When we average over the azimuthal angles of the produced hadrons,
only the diagonal elements of the density matrix, as well as the $\ell
0$ interference term survive in the cross section. They were
calculated in ref.~\cite{GKP} and we give them again for completion:
\bea\label{eq2.7}
L_{RR} &=& \frac{Q^2}{|\vec q|^2} (2E-\nu+|\vec q|)^2
-\frac{m_\mu^2}{|\vec q|^2} \left[2\nu(2E-\nu+|\vec q|)+m_\mu^2\right]\nonumber\\
L_{LL} &=& \frac{Q^2}{|\vec q|^2} (2E-\nu-|\vec q|)^2
-\frac{m_\mu^2}{|\vec q|^2} \left[2\nu(2E-\nu-|\vec q|)+m_\mu^2\right]\nonumber\\
L_{00} &=& \frac{2\,[Q^2(2E-\nu)-\nu m_{\mu}^2]^2}{Q^2 |\vec q|^2} - 2(Q^2+m_\mu^2)\\
L_{\ell\ell} &=& 2 m_{\mu}^2
\left(\frac{m_{\mu}^2}{Q^2}+1\right)\nonumber\\
L_{\ell 0} &=& \frac{2m_\mu^2[Q^2(2E-\nu)-\nu m_{\mu}^2]}{Q^2
|\vec q|}\nonumber
\eea
All matrix elements are positive in the physical region, which becomes
evident when the kinematic condition $Q_{min}^2=m_{\mu}^2
\frac{\nu}{E-\nu}$ is taken into account. For the propagator of the
spin-$3/2$ resonance we introduce the Rarita-Schwinger propagator in
free space~\cite{korpa}
\bea
G_{\mu\nu}(p_{\Delta}) &=&
\frac{\pslash_{\Delta}+M_{\Delta}}{p_{\Delta}^2-M_{\Delta}^2+iM_{\Delta}\Gamma_{\Delta}}
\left[
g_{\mu\nu}-\frac{1}{3}\gamma_{\mu}\gamma_{\nu}-\frac{2}{3}\frac{1}{M_{\Delta}^2}p_{\Delta\mu}p_{\Delta\nu}
+\frac{1}{3}\frac{1}{M_{\Delta}}(p_{{\Delta}\mu}\gamma_{\nu} -
p_{{\Delta}\nu}\gamma_{\mu}) \right]\nonumber\\
&\equiv&
\frac{\pslash_{\Delta}+M_{\Delta}}{p_{\Delta}^2-M_{\Delta}^2+iM_{\Delta}\Gamma_{\Delta}}\,G'_{\mu\nu}
\eea
which is sufficient for the present work. The mass of
$M_{\Delta}=1232$~MeV and the width $\Gamma_{\Delta}=120$~MeV will be
taken as constants. But in some articles it is a function of the
invariant mass $\Gamma_{\Delta}=\Gamma_{\Delta 0}\left( \frac
{p_{\pi}(W)} {p_{\pi}(M_{\Delta})}\right)^3$ with
$p_{\pi}(W)=\frac{1}{2 M_{\Delta}} \sqrt{(W^2-M_N^2-m_{\pi}^2)^2-4
M_N^2 m_{\pi}^2}$. Several authors studied the modifications of the
propagator in nuclear matter, which will be useful when we consider
nuclear corrections.\\

The matrix element for the entire process includes the coupling of
the $Wp\Delta$-vertex given in terms of form factors and the $\pi
p \Delta$ coupling $ig_{\Delta} p_{\pi}^{\mu}$. The matrix element
is
\beq
{\cal M}_h = {\sqrt 3} \; \; ig_{_\Delta} p_{\pi}^{\mu}\;\bar u(p')
 \left[G_{\mu\nu}(p+q)\; d^{\nu\lambda}
        +\frac{1}{3} G_{\nu\mu}(p-p_{\pi}) \;d^{\lambda\nu}
\right] u(p)\;\varepsilon_{\lambda}(q,h).
\eeq
For our process there are two propagators; one in the $s$-channel for
$\Delta^{++}$, as shown in figure 1, and one in the $u$-channel for
$\Delta^0$. The arguments of the propagator are $(p+q)^2$ and
$(p-p_{\pi})^2$, respectively. The argument of the polarization is
$q_{\mu}$ and $h$ denotes helicity. In this preprint we include, for
the calculation and the curves, only the $s$-channel pole which
resonates. When we submit the article for publication both terms will
be included.\\

The coupling at the $Wp\Delta$-vertex are included in $d^{\nu\lambda}$
which will be discussed below. The $\pi p \Delta$ coupling is taken
from Appendix A1 of ref.~\cite{LPP}:
$g_{_\Delta}=15.3$~GeV$^{-1}$. The factor $\sqrt{3}$ originates from
the isospin relation
\beq \langle \Delta^{++}|V_{\mu}^+|p\rangle = \sqrt 3 \langle
\Delta^{+}|V_{\mu}^3|p\rangle
\eeq
where the right hand side is related to the electromagnetic form
factor, whose numerical value was determined in early experiments and
are used in many articles. Electroproduction data have been used as an
input for neutrino reactions\footnote{To our knowledge the factor
$\sqrt 3$ was introduced first by Schreiner and von
Hippel~\cite{schreiner} and has become traditional to keep it in recent
calculations.} and this convention still survives. The factor $1/3$ in
front of the $u$-channel pole comes from the Clebsch-Gordan
coefficients. \\

The $Wp\Delta$-vertex contains vector and axial form factor included
in the function
\bea
d^{\nu\lambda} &=& g^{\nu\lambda} \left[ \frac{C_3^V}{M_N} \qslash
+\frac{C_4^V}{M_N^2} p_{\Delta}.q +\frac{C_5^V}{M_N^2} p.q + C_6^V
\right] \gamma_5
-q^{\nu} \left[ \frac{C_3^V}{M_N} \gamma^{\lambda}
+\frac{C_4^V}{M_N^2} p_{\Delta}^{\lambda} +\frac{C_5^V}{M_N^2} p^{\lambda}
\right] \gamma_5\nonumber\\
&&+g^{\nu\lambda}C_5^A + q^{\nu}q^{\lambda}\frac{C_6^A}{M_N^2}.
\eea
The vector form factors were determined~\cite{LPP} using
electroproduction data. Among the axial form factor the most
important are $C_5^A(q^2)$ and $C_6^A(q^2)$ and for this reason we
omitted the other two axial form factors. All form factors will be
given explicitly in the next section. The square of the matrix
element for the $s$-channel pole is
\bea
{\cal M}^{h' *}{\cal M}^{h} = \frac{3}{2}
\frac{g_{\Delta}^2}{(p_{\Delta}^2-M_{\Delta}^2)^2+M_{\Delta}^2
\Gamma_{\Delta}^2}
Tr[&&p_{\pi}^{\mu} (\pslash_{\Delta}+M_{\Delta}) G'_{\mu\nu}
d^{\nu\lambda} \varepsilon_{\lambda}(h) (\pslash+M_N)
\nonumber\\&&\varepsilon^*_{\lp}(h') d^{\nu^\prime\lp}
G'_{\mu^\prime\nu^\prime} (\pslash_{\Delta}+M_{\Delta})
p_{\pi}^{\mu^\prime} (\pslash^\prime+M_N)].
\eea
The factor $1/2$ comes from averaging over initial spins of the
target. The helicity cross sections and interference terms for the
scattering of the current on a proton target are defined as
\beq\label{eq2.13}
\frac{d\sigma^{h'h}}{dE_{\pi}}(\nu,Q^2) = \frac{1}{32\pi\nu
M_N|{\vec p}_{\Delta}|} {\cal M}^{h' *}{\cal M}^{h}
\eeq
We now have all the ingredients for calculating helicity cross
sections for the processes $W^+p\ra {\cal R}^{++}\ra\pi^+ p$. The
calculation is straight forward since it involves a two-body phase
space and a trace. It is long because of the many
$\gamma$-matrices occurring in the propagator and the
$Wp\Delta$-vertex. The trace calculation was done using
FEYNCALC~\cite{Feyncalc}. \\

Finally we can include the lepton variables and present the triple
differential cross section
\beq \frac{d\sigma}{dE_{\pi}\,dQ^2\,d\nu} = \frac{G^2}{4\pi^2}
\frac{\nu}{4E^2} |V_{ud}|^2 \left[ L_{00} \frac{d\sigma^S}{dE_{\pi}} + L_{LL}
\frac{d\sigma^L}{dE_{\pi}} + L_{RR} \frac{d\sigma^R}{dE_{\pi}}
+L_{\ell\ell} \frac{d\sigma^{\ell}}{dE_{\pi}} + 2\,L_{\ell 0}
\frac{d\sigma^{\ell 0}}{dE_{\pi}}\right].
\eeq
This formula includes the muon mass contained in the $L_{h'h}$
functions. In the limit $m_{\mu}=0$ it reduces to the known
result~\cite{BjPaschos}. We note that the formalism simplifies the
calculations because the leptonic part was incorporated as an
overall factor. We also note that there is only one interference
term $\frac{d \sigma^{\ell 0}}{d E_{\pi}}$ because the other
interference terms vanish when we average over the azimuthal angle
of the produced hadrons.

\section{Numerical Estimates}
\label{sec:Numerical}

Besides the form factors we have now all the ingredients for
calculating the pion spectrum. The vector form factors have been studied
in earlier paper determining their $Q^2$-dependence from
electroproduction data~\cite{LPP}. It has been established that they are
modified dipoles
\beq
C_3^V(Q^2) = \frac{C_3^V(0)}{(1+Q^2/M_V^2)^2}
\frac{1}{1+Q^2/(4\,M_V^2)}
\eeq
with $C_3^V(Q^2=0)=1.95$ and $M_V=0.84$~GeV. The dominance of the
magnetic dipole gives the relation
\beq
C_4^V(Q^2) = -C_3^V(Q^2)\, \frac{M_N}{W}, \quad C_5^V = 0.
\eeq
The other two terms $C_5^V=C_6^V$ were set to zero. These form factors
were determined by electroproduction data where it was shown that they
reproduce the measured helicity amplitudes~\cite{LPP}. 

The axial couplings were obtained from the decay rate of the
$\Delta$-resonance and the reproduction of neutrino data~\cite{LPP}  
\beq
C_5^A(Q^2) = \frac{C_5^A(0)}{(1+Q^2/M_A^2)^2}
\frac{1}{1+Q^2/(3\,M_A^2)}
\eeq
with $C_5^A(Q^2=0)=1.2$ and $M_A=1.05$~GeV. PCAC gave us
\beq
C_6^A(Q^2) = C_5^A(Q^2) \frac{M_N^2}{Q^2+m_{\pi}^2}.
\eeq
$C_6^A(Q^2)$ is the pseudoscalar form factor with its contribution
to the cross section being proportional to the square of the
lepton mass. \\

\begin{figure}[htb]
\vspace*{1.5cm}
\centerline{ \epsfxsize=12.0cm\epsfysize=12.0cm
\epsfbox{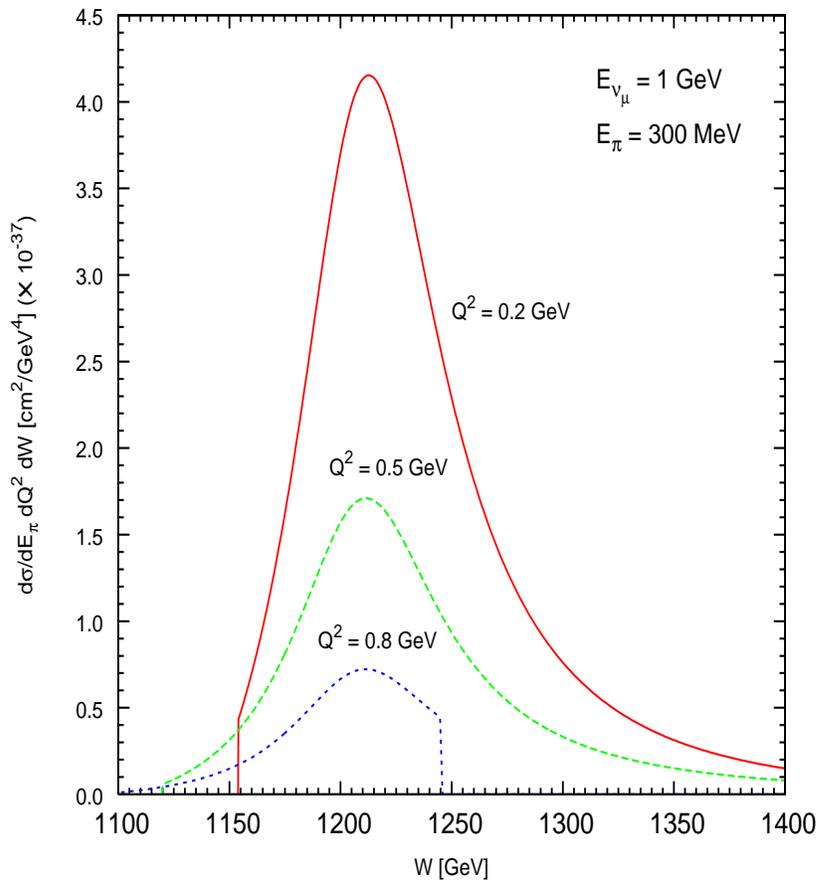} } \caption{\em Triple differential cross section
for a $1$~GeV $\nu_{\mu}$ interacting with a proton for different
$Q^2$. 
         } \label{figure02}
\vspace*{-0.2cm}
\end{figure}

\begin{figure}[htb]
\vspace*{1.5cm}
\centerline{ \epsfxsize=12.0cm\epsfysize=12.0cm
\epsfbox{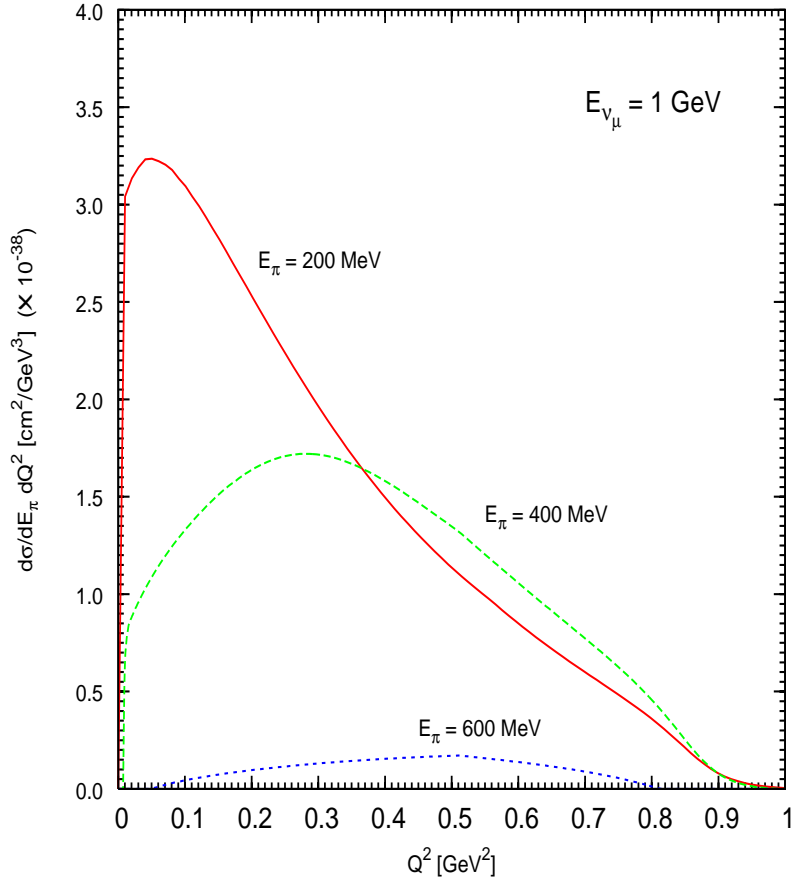} } \caption{\em Double differential cross section for a $1$~GeV $\nu_{\mu}$ interacting with a proton for different
$E_{\pi}$. This is calculated from Fig.\ref{figure02} after $W$
integration.  } \label{figure03}
\vspace*{-0.2cm}
\end{figure}

The triple differential cross section
$\frac{d\sigma}{dE_{\pi}\,dQ^2\,dW}$ shown in figure~\ref{figure02}
for neutrino energy of $1.0$~GeV, $E_{\pi}=300$~MeV, $Q^2=0.2,0.5$ and
$0.8$~GeV$^2$. The curves show a $\Delta$-peak which is a sensitive
function of $Q^2$. This is expected since for large values of $Q^2$
there is the large decrease of the form factors. Integrating over $W$
one obtains the double differential cross section of
figure~\ref{figure03}. Again the cross section decrease with
increasing $E_{\pi}$ because the process runs out of phase space. \\

Finally more interesting is the dependence on the pion energy when
all other variables are integrated. Figure~\ref{figure04} shows the
contribution of the important form factors. the term $C_5^A$ dominates
with the next contribution coming from $C_3^V$. The interference
between $C_3^V$ and $C_4^V$ is destructive. The interference between
vector and axial form factors is constructive for neutrinos and
destructive for anti-neutrinos. In figure~\ref{figure05} we set the
muon mass equal to zero in order to see the effect of neglecting the
mass.  We repeat the calculations for higher neutrino energies
$E_{\nu_{\mu}}= 1,2$, and $5$~GeV shown in
figures~\ref{figure06}-\ref{figure08}. \\

The mass of the charged lepton influences the pion spectra shown in
Figures~\ref{figure04} and \ref{figure05}. Mass effects are much more
prominent for $\nu_{\tau}$ beams, where the threshold effect is
dominant. In Fig.~\ref{figure09} we show the pion spectrum for
$E_{\nu_\tau} = 5$~GeV.\\

A new feature is the change in the significance of the various form
factors. The induced pseudoscalar form factor $C_6^A$ is more
important relative to $C_5^A$. The integrated cross section is smaller
because it reaches its asymptotic value at a much higher energy. In
Fig.~\ref{figure10} we show the integrated cross section as a function
of $E_{\nu_\tau}$. In all the figures, we have included the spectra
for anti-neutrinos, which are obtained by changing the sign of the
vector $\otimes$ axial interference terms. In closing, we remark that the pion
spectra show relevant new features, and will be important in
deciphering the importance and the functional dependence of the form
factors.\\

For comparison with other articles one must keep in mind that we did
not include nuclear effects from the target. We decided to use protons
or neutrons as free targets in order to study the significance of the
various form factors. We may include nuclear target effects later on.\\

Several articles calculated and presented the pion spectrum and we
comment on them. An early article~\cite{PPY}, where one of us is a
co-author, presented in figures 8-16 pion energy spectra with a
different shape because the phase space was treated incorrectly. The
discrepancy was noticed and corrected in figures 3 and 4 of
ref.~\cite{PSU}. The same discrepancy has been pointed out when the
spectrum was calculated in~\cite{siegen}. Between the previous two
articles, mentioned above, and the present article there is a
difference in the method of calculation.  The earlier articles
calculated the production of the delta resonance and then folded its
decay into a pion and a nucleon.  This method however requires
knowledge of the density matrix elements as described in Eqs. (1.3)
and (1.4) of ref.~\cite{schreiner}, because resonances in various
polarization states produce pions with different energies.  In the
present article we calculate the entire process with the
$\Delta$-resonance in the intermediate state. The same procedure is
advocated in a recent article~\cite{athar}. Thus the pion energy
spectrum on a Hydrogen target is a rather interesting quantity being
sensitive to the form factors.\\

\section{Summary}
\label{sec:Summary}

Neutrino interactions are reaching an age of theoretical maturity and
will come to be compared with the new generation of experiments. For
this reason we investigated the neutrino cross sections in the energy
range of the $\Delta$-resonance. To this end we have rewritten the
neutrino-nucleon cross section in terms of helicity of cross sections
of the $W^\pm$ on nucleons. For the sake of brevity, we have not
included neutral current reactions, which will be presented in the
future. Our plan is to present these results in a code and explicit
publications.\\

\begin{figure}[htb]
\vspace*{1.5cm}
\centerline{ \epsfxsize=12.0cm\epsfysize=12.0cm
\epsfbox{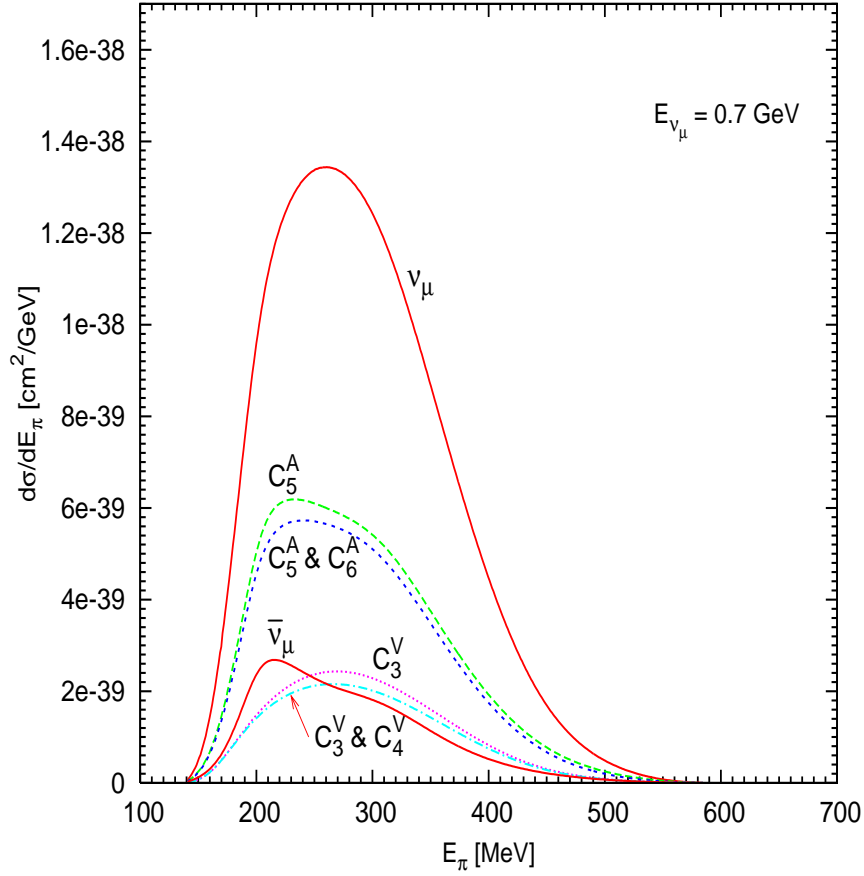} } \caption{\em Pion energy spectrum for an
incoming $\nu_{\mu}$ of energy $0.7$~GeV. Contributions from several
from factors are shown separately. Here, for example, $C_5^A\& C_6^A$
means in this case we have put only these two form factors
finite. Hence it contains their interference term as well. By
$\nu_{\mu}$ and $\bar\nu_{\mu}$ we denote the contributions of all the
form factors to cross sections for these particles. Here muon mass
effects are taken into account. For $\nu_{\mu}$ a constructive
interference is observed, while for $\bar\nu_{\mu}$ it is destructive.
} \label{figure04}
\vspace*{-0.2cm}
\end{figure}

\begin{figure}[htb]
\vspace*{1.5cm}
\centerline{ \epsfxsize=12.0cm\epsfysize=12.0cm
\epsfbox{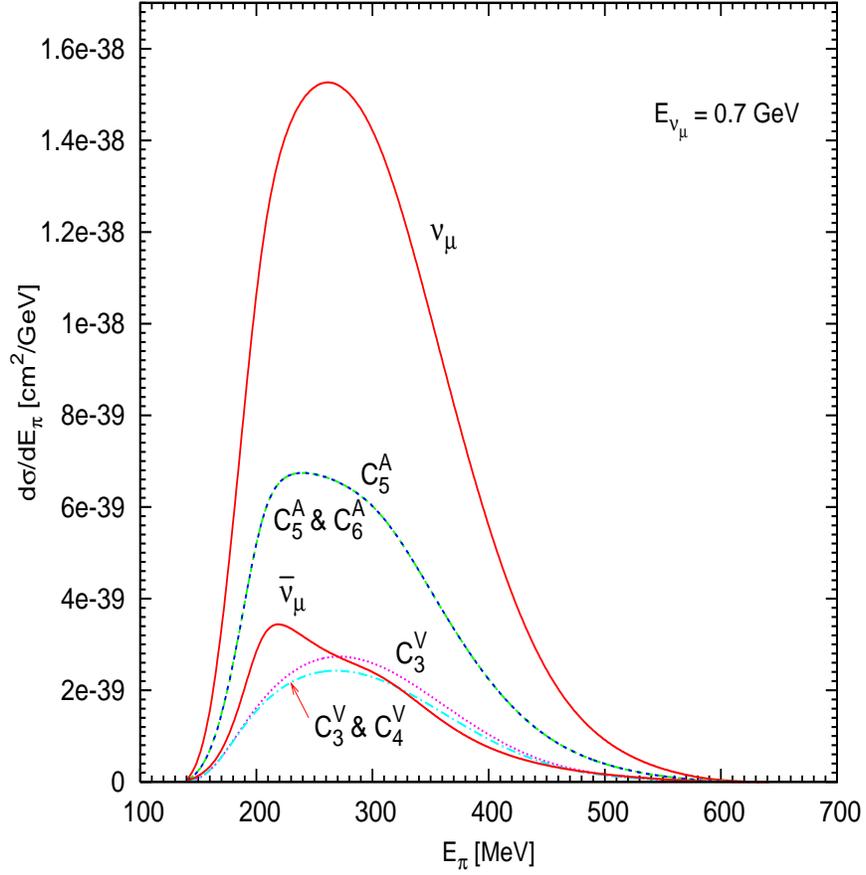} } \caption{\em Same as
Fig.~\ref{figure04} but with outgoing muon mass neglected. In this
limit $C_6^A$ does not contribute. These cross sections are a bit
enhanced compared to Fig.~\ref{figure04} due to more phase space in
this limit. 
         } \label{figure05}
\vspace*{-0.2cm}
\end{figure}

\begin{figure}[htb]
\vspace*{1.5cm}
\centerline{ \epsfxsize=12.0cm\epsfysize=12.0cm
\epsfbox{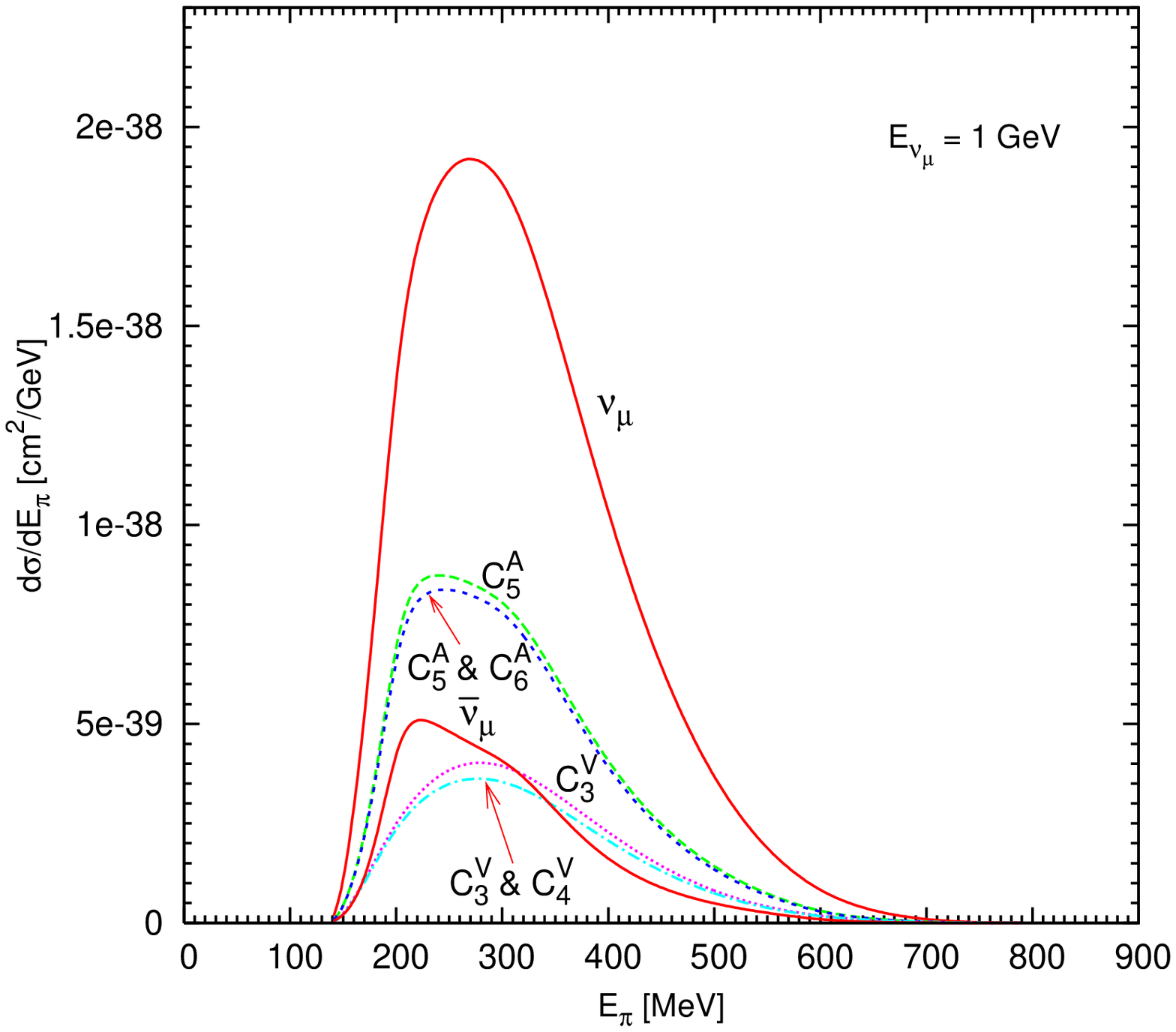} } \caption{\em Pion energy spectrum for an
incoming $\nu_{\mu}$ of energy $1$~GeV. Notations are similar to 
Fig.~\ref{figure04}. 
         } \label{figure06}
\vspace*{-0.2cm}
\end{figure}

\begin{figure}[htb]
\vspace*{1.5cm}
\centerline{ \epsfxsize=12.0cm\epsfysize=12.0cm
\epsfbox{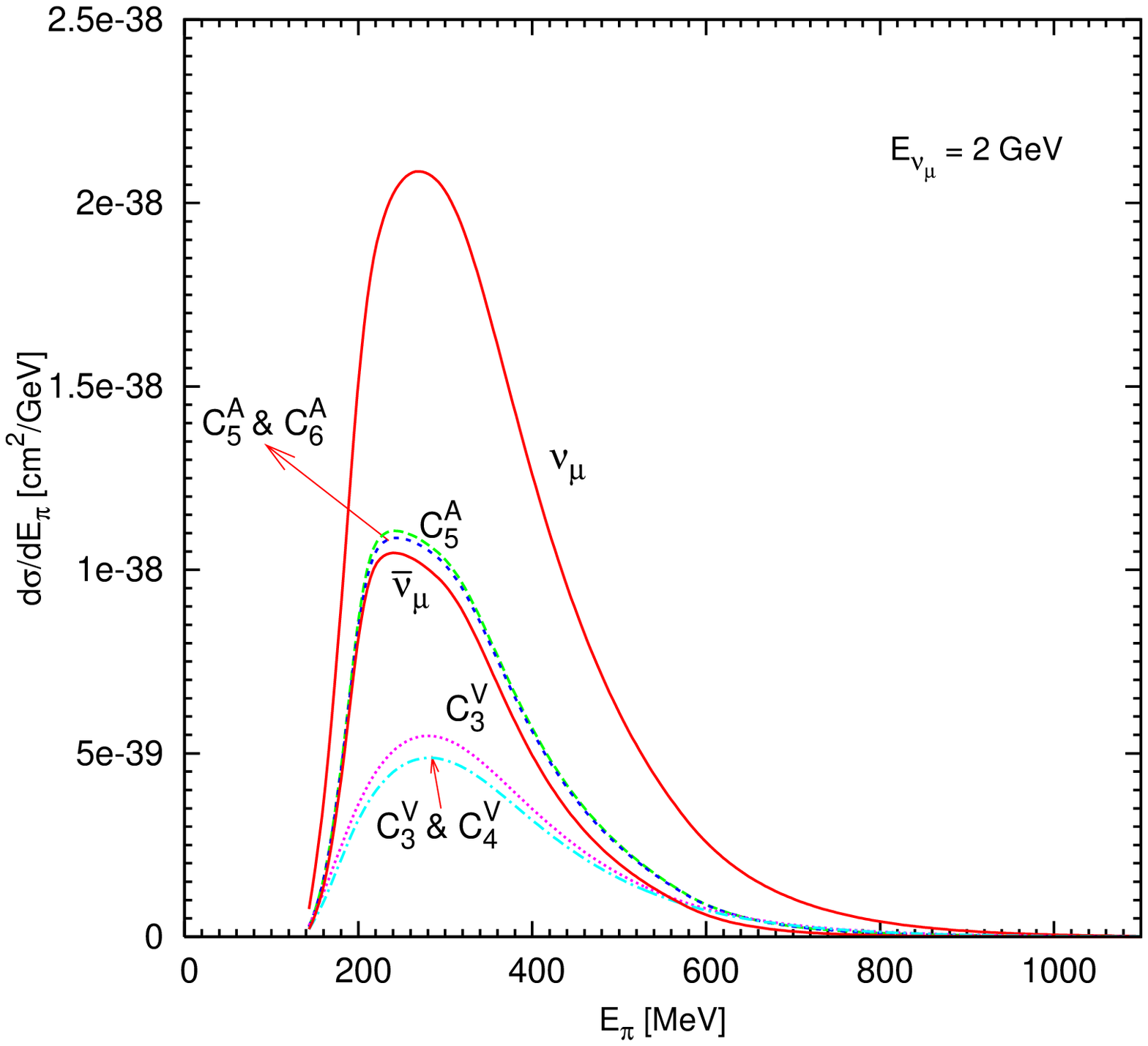} } \caption{\em Pion energy spectrum for an
incoming $\nu_{\mu}$ of energy $2$~GeV. Notations are similar to 
Fig.~\ref{figure04}. 
         } \label{figure07}
\vspace*{-0.2cm}
\end{figure}

\begin{figure}[htb]
\vspace*{1.5cm}
\centerline{ \epsfxsize=12.0cm\epsfysize=12.0cm
\epsfbox{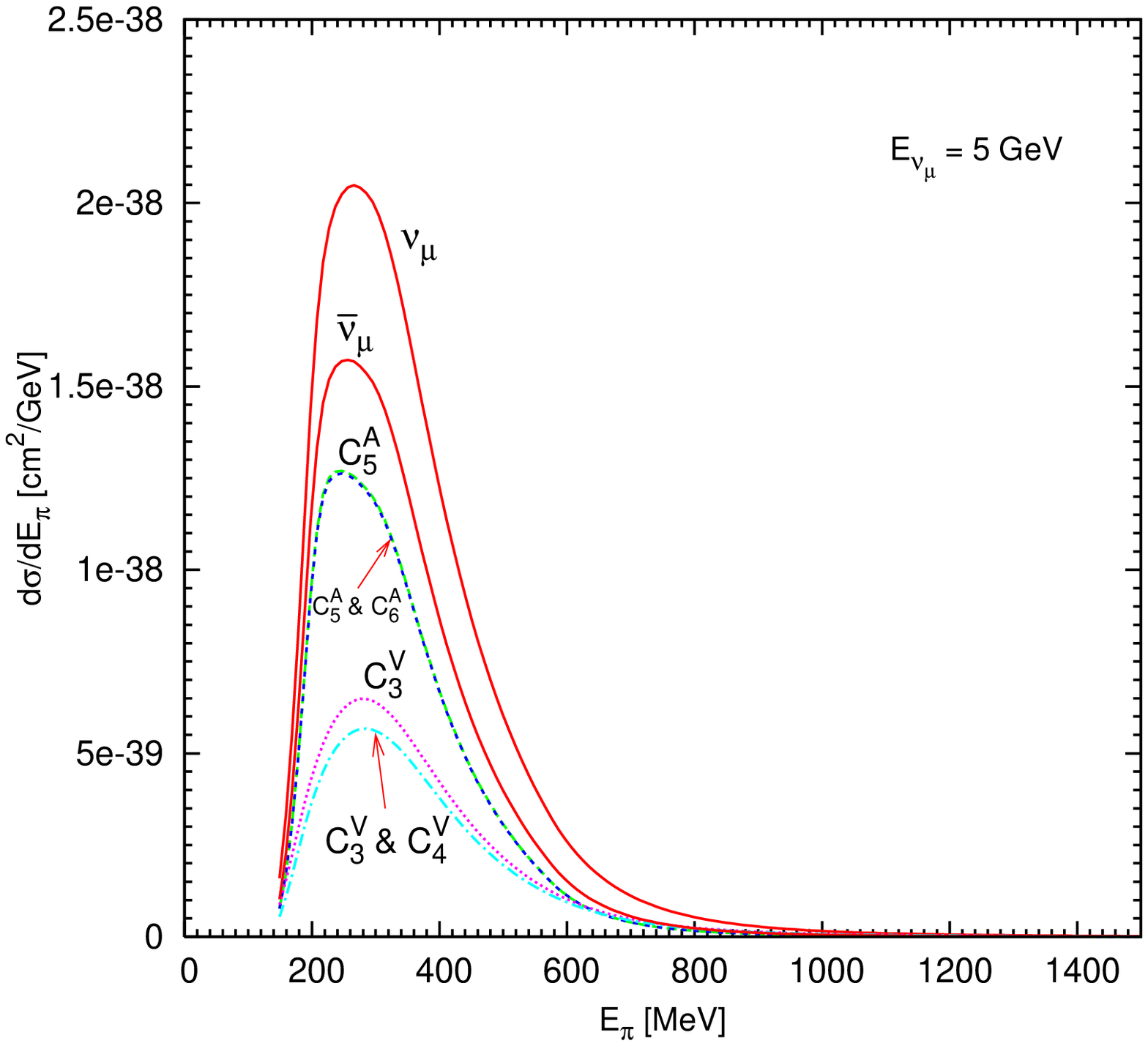} } \caption{\em Pion energy spectrum for an
incoming $\nu_{\mu}$ of energy $5$~GeV. Notations are similar to 
Fig.~\ref{figure04}. 
         } \label{figure08}
\vspace*{-0.2cm}
\end{figure}

\begin{figure}[htb]
\vspace*{1.5cm}
\centerline{ \epsfxsize=12.0cm\epsfysize=12.0cm
\epsfbox{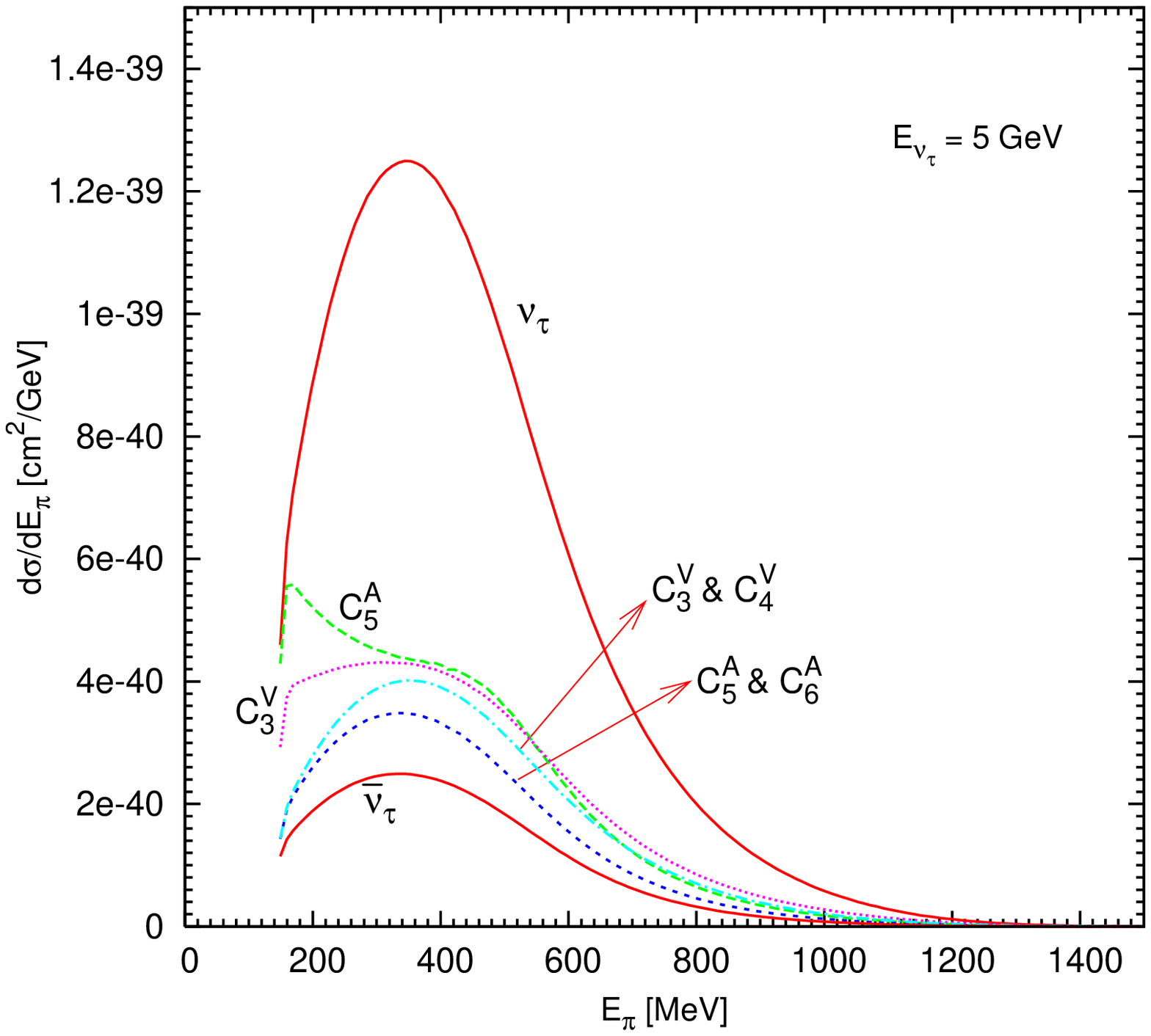} } \caption{\em Pion energy spectrum for an
incoming $\nu_{\tau}$ of energy $5$~GeV. Notations are similar to 
Fig.~\ref{figure04}. 
         } \label{figure09}
\vspace*{-0.2cm}
\end{figure}

\begin{figure}[htb]
\vspace*{1.5cm}
\centerline{ \epsfxsize=12.0cm\epsfysize=12.0cm
\epsfbox{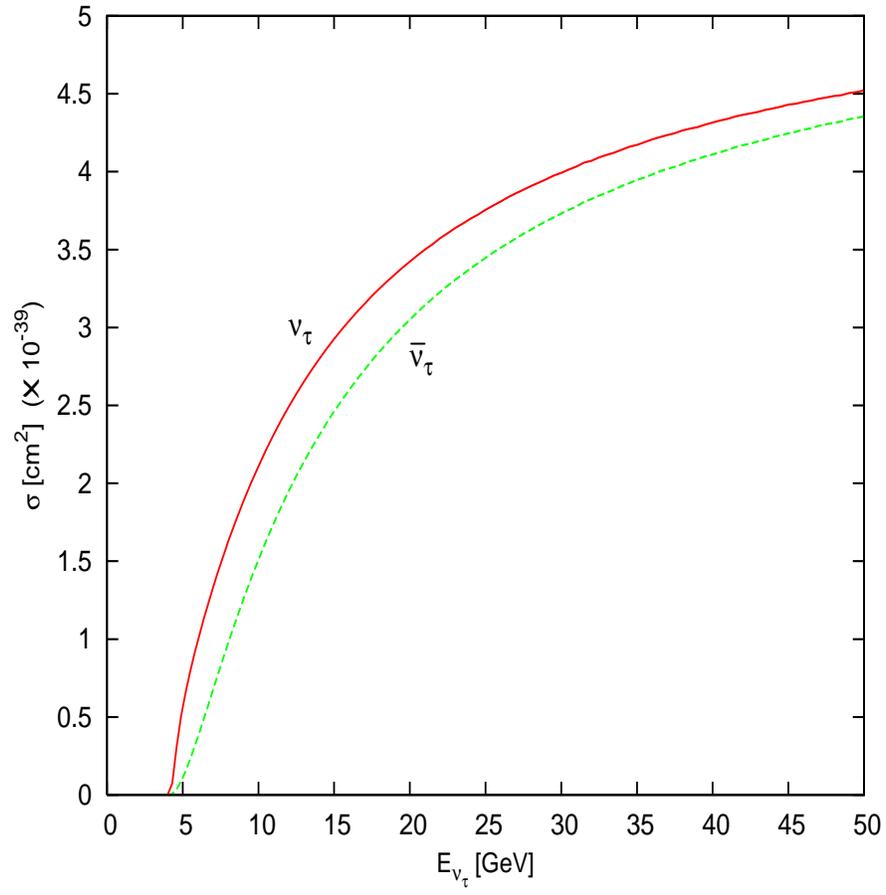} } \caption{\em Variation of $\nu_{\tau}$ total
cross section with neutrino energy. 
         } \label{figure10}
\vspace*{-0.2cm}
\end{figure}

Besides the formalism, we have made the following improvements:

\begin{enumerate}

\item We included the charged lepton mass. The reader who wishes to see
effects originating form the mass can set them equal to zero in
Eqs.~\ref{eq2.7} and include them in the phase space of the two-body
cross section given in Eq.~\ref{eq2.13}. Such a comparison was presented in
Figs.~\ref{figure04} and \ref{figure05}.

\item We use a running width for the resonance as described after
Eq.~\ref{eq2.13}. This was also included in ref.~\cite{LPP}.

\item We study the significance of the pseudo-scalar form factor $C_6^A$ 
for muon and tau neutrino-induced reactions. We have also presented
contributions from the various form factors.
\end{enumerate}

With this article, we hope to clarify several questions presented by
colleagues who are planning and carrying out the experiments. There
are several other quantities that need to be calculated, and for just
this reason, we have developed a flexible formulation which can be
adapted to new situations which may arise. Finally, as previously
mentioned, we are preparing a CODE which will cover new demands that
may come up in the future.

\acknowledgments
One of us (SR) is thankful to `Bundesministerium f\"ur Bildung und
Forschung', Berlin/Bonn for financial support and to Prof.~E.~Reya for
providing the research infrastructure during his stay at Technische
Universit\"{a}t Dortmund. The other one (EAP) acknowledges the early
collaboration with Dr.~J.~Y.~Yu on topics related to this article.


\newpage
\end{document}